\documentclass[12pt,a4paper,twoside]{article}

\usepackage{graphicx}
\usepackage{fancyhdr,calc,cite,float}
\usepackage{epsfig}
\usepackage[centertags]{amsmath}
\usepackage[mathscr]{eucal}
\usepackage[T1]{fontenc}
\usepackage{amssymb}
\newcommand{\lambdabar}{\ensuremath{\raisebox{-1pt}{$\mathchar'26\mkern-10mu$}\lambda}}

\title{\bf Excitation of heavy hydrogen-like ions by light atoms  
in relativistic collisions with large momentum transfers} 

\author{\bf B. Najjari and A.B. Voitkiv  \\ 
Max-Planck-Institut f\"ur Kernphysik \\ 
Saupfercheckweg 1, 69117 Heidelberg, Germany }

\date{\today}  

\begin{document}

\maketitle

\begin{abstract} 

We present a theory for  
excitation of heavy hydrogen-like projectile-ions  
by light target-atoms in collisions where  
the momentum transfers to the atom  
are very large on the atomic scale.  
It is shown that in this process 
the electrons and the nucleus of the atom 
behave as (quasi-) free particles 
with respect to each other and that 
their motion is governed by the field 
of the nucleus of the ion. 
The effect of this field on the atomic particles  
can be crucial for the contribution to the excitation 
of the ion caused by the electrons of the atom.  
Due to comparatively very large nuclear mass, 
however, this field can be neglected 
in the calculation of the contribution 
to the excitation due to the nucleus of the atom.  
  
\end{abstract}



\section{Introduction} 

Projectile-electron excitation and loss 
occurring in collisions of projectile-ions 
with atomic targets in the non-relativistic 
domain of impact energies and projectile charges 
have been extensively studied 
during the last few decades  
(for a review see \cite{MMM}-\cite{SDR}). 
Starting with pioneering articles 
of Bates and Griffing \cite{B-G}   
the most of the theoretical studies of 
these processes have been based on the first order
perturbation theory in the projectile-target 
interaction \cite{mcguire-book}.  

In contrast to nonrelativistic collisions, 
in the relativistic domain of collision parameters 
the first consistent theoretical approaches for treating 
projectile-electron excitation and loss was formulated 
relatively recently \cite{voit-1999}. 
Like in the nonrelativistic domain, 
the simplest description of these processes  
is given by the first order theory in the interaction  
between the projectile and the target. 
This theory is strictly valid provided the following 
conditions are fulfilled simultaneously: 
i) $Z_I \ll v$ and ii) $Z_A \ll v$, where 
$Z_I$ and $Z_A$ are the atomic numbers of 
the projectile and target, respectively, and  
$v$ is the collision velocity 
(atomic units are used throughout except where 
otherwise stated). 

In the present article we shall discuss 
excitation of heavy hydrogen-like projectiles 
in collisions with very light atoms when 
the condition $Z_A \ll v$ is very well fulfilled but 
the atomic number of the projectile is so high that 
one has $Z_I \simeq v$ even for collision velocities from 
the relativistic domain $v \sim c$, where $c$ is the speed of light.  
In such a case the field of the target {\it per se} 
represents just a weak perturbation for the electron of the projectile. 
Nevertheless, large higher order contributions to the projectile-electron 
excitation may arise "indirectly" since the strong field 
of the nucleus of the projectile 
can substantially distort the motion of the target electrons 
which in turn changes the cross section for excitation.   

Note that some aspects of projectile-electron 
excitation and loss in asymmetric collisions ($Z_I \gg Z_A$) 
have been already considered in \cite{voit-najj-nreik} and \cite{voit-reik}  
where eikonal-like theories of these processes were formulated for  
the nonrelativistic \cite{voit-najj-nreik}  
and relativistic \cite{voit-reik} domains.  
Those theories, however, were based on the assumption 
that both in the initial and 
final channels the motion of the electrons of the target  
is mainly driven by the field of the atomic nucleus while  
the field of the projectile just distorts somewhat this motion. 
In the present article, the work on which was triggered by a recent experiment 
performed at GSI (Darmstadt, Germany) \cite{gsi-experiment}, 
we consider the opposite case which is realized 
in collisions characterized by very large (on the target scale) 
momentum transfers. It turns out that under such condition 
it is the field of the nucleus of the projectile 
which is the main driving force in the collision 
not only for the electron of the projectile 
but also for those of the target.      

The article is organized as follows.    
In the next section, based on the first order theory 
in the projectile-target interaction, 
we show that in collisions with large momentum transfers 
the electrons and the nucleus of the target atom 
act incoherently behaving like quasi-free particles 
with respect to each other. 
In this section we also argue that the motion 
of the atomic particles in such collisions 
is driven primarily by the field of 
the nucleus of the ion and, therefore, 
a better (compared to the first order)
treatment of the excitation can be obtained 
by describing the electrons of the atom 
in their initial and final states 
as moving in the field of the nucleus of the ion.    
In sections 3 and 4 we present treatments   
for the excitation by proton and electron impacts, 
respectively. In section 5 the theory is illustrated 
by calculating cross sections for excitation of 
Bi$^{82}(1s)$ ions in relativistic collisions with hydrogen. 

\section{ Excitation in collisions with light atoms }

Let us consider excitation of a heavy hydrogen-like ion 
in collisions with light atoms. For the moment we shall assume, 
for the sake of simplicity, that the atom consists 
of a nucleus with a charge $Z_A$ and just one electron.  

It can be shown (see e.g. \cite{voitkiv-buch}) that within 
the first order approximation in the ion-atom interaction 
the cross section for the ion-atom collision reads 
\begin{eqnarray} 
\frac{d^2 \sigma_{0 \rightarrow n}^{0 \rightarrow m} }%
{d^2 {\bf q}_{\perp}} = \frac{4}{v^2} %
\frac{\left| \left(F^I_0 + \frac{v}{c} F^I_3 \right) %
\left(F^0_A + \frac{v}{c} F^3_A \right) + %
\frac {F^I_3 F^3_A}{\gamma^2} +%
\frac {F^I_1 F^1_A +F^I_2 F^2_A }{\gamma} \right|^2}%
{\left({\bf q}_i^2 -   
\frac{(\varepsilon_n-\varepsilon_0)^2}{c^2} \right)^2 }.   
\label{a1} 
\end{eqnarray} 
Here, $\varepsilon_0$ and $\varepsilon_n$  
are initial and final internal energies
of the ion, respectively  
and $ {\bf q}_i$ is the 3-momentum transferred to 
the ion; all the quantities are given in the rest frame of the ion.   
We also introduce quantities $\epsilon_0$, $\epsilon_m$ and 
$ {\bf q}_a$ which have similar meanings but
are for the atom and given in the rest frame of the atom. 
The momentum transfers are defined by 
$ {\bf q}_i = ({\bf q}_{\perp}, q_{min}^i) $ and 
$ {\bf q}_a = (-{\bf q}_{\perp}, -q_{min}^a) $, where
${\bf q}_{\perp}$ is the two-dimensional part of the
momentum transferred to the atom, which is perpendicular
to the collision velocity ${\bf v}$, and the 
components of the momentum transfers
along the collision velocity read 
\begin{eqnarray}
q_{min}^i &=& \frac{\varepsilon_n - \varepsilon_0}{v} + %
\frac{\epsilon_m-\epsilon_0}{v \gamma} %
\nonumber \\
q_{min}^a & = & \frac{ \epsilon_m - \epsilon_0}{v} + %
\frac{\varepsilon_n-\varepsilon_0}{v\gamma},  
\label{a2}
\end{eqnarray} 
where $\gamma = 1/\sqrt{1- v^2/c^2}$ is the collisional Lorentz factor.  
The inelastic 4-component form-factor 
of the ion (in the ion frame) and the 4-component form-factor 
of the atom (in the atom frame) are given by 
\begin{eqnarray}
&& F^I_{0} \equiv F^I_{0}({\bf q}_i ) %
= - \langle \varphi_n | %
\exp(i {\bf q}_i \cdot {\bf r}) | %
\varphi_0  \rangle 
\nonumber \\ 
&& F^I_{l} \equiv F^I_{l}({\bf q}_i ) %
= \langle \varphi_n |   %
\exp(i {\bf q}_i \cdot {\bf r}) %
\, \alpha_{l} | \varphi_0 \rangle    
\label{a3} 
\end{eqnarray}
\begin{eqnarray}
F_{A}^{0} \equiv F_{A}^{0}({\bf q}_a ) &=& %
Z_A \delta_{m0} - \langle u_m | \exp(i {\bf q}_a \cdot %
\mbox{\boldmath$\xi$}) | u_0 \rangle 
\nonumber \\
F_{A}^{l} \equiv F_{A}^{l}({\bf q}_a ) &=&  %
- \langle u_m | \alpha_l \, %
\exp(i {\bf q}_a \cdot \mbox{\boldmath$\xi$}) | u_0 \rangle,  %
\label{a4}
\end{eqnarray}
respectively ($l=1,2,3$). In Eq.(\ref{a3}) 
$\varphi_0=\varphi_0({\bf r})$ and $\varphi_n=\varphi_n({\bf r})$
are the initial and final
internal states of the ion, ${\bf r}$
the coordinates of the ion electron with respect to
the ion nucleus and $\alpha_l$ 
the Dirac matrices for the electron of the ion. 
In Eq.(\ref{a4})
$u_0=u_0(\mbox{\boldmath$\xi$})$ 
and $u_m=u_m(\mbox{\boldmath$\xi$})$
are the initial and final internal states of the atom, 
and $\mbox{\boldmath$\xi$}$ are the coordinates of  
the atomic electron with respect to the atomic
nucleus. 

If one is interested only in the electron transitions 
in the ion, without knowing what occurs with the atom in the collision,  
then one has to consider the cross section 
\begin{eqnarray} 
\frac{d^2 \sigma_{0 \rightarrow n} }{d^2 {\bf q}_{\perp}} = 
\sum_{m} \frac{d^2 \sigma_{0 \rightarrow n}^{0 \rightarrow m} }  
{d^2 {\bf q}_{\perp}}, 
\label{a5} 
\end{eqnarray} 
where the sum runs over all possible internal states
of the atom including its initial state 
and the atomic continuum. 
The cross section (\ref{a5}) can be conveniently written as 
the sum of the contributions from the elastic ($m=0$) and inelastic 
($m \neq 0$) atomic collision modes     
\begin{eqnarray} 
\frac{d^2 \sigma_{0 \rightarrow n} }{d^2 {\bf q}_{\perp}} = 
\frac{d^2 \sigma_{0 \rightarrow n}^{0 \rightarrow 0} }{d^2 {\bf q}_{\perp}} + 
\sum_{m \neq 0} \frac{d^2 \sigma_{0 \rightarrow n}^{0 \rightarrow m} }
{d^2 {\bf q}_{\perp}}, %
\label{a6} 
\end{eqnarray} 
where the sum over $m \neq 0 $, 
which represents the contribution from the inelastic atomic mode, 
includes also the sum over the continuum states of the atom. 

In collisions resulting in excitation of heavy hydrogen-like ions 
the momentum transfer to the atom $q_a$ can, under certain conditions, be 
much larger than the typical momentum ($\simeq Z_A$) 
of the electron bound in the atom. From the second equation in 
(\ref{a2}) one may see that this will be certainly 
the case when $(\varepsilon_n-\varepsilon_0)/(v\gamma) \gg Z_A$.   
Taking into account that $\varepsilon_n-\varepsilon_0 \sim Z_I^2$ 
we obtain that provided the condition 
\begin{eqnarray} 
\frac{ Z_I^2 }{ Z_A } \gg v\gamma   
\label{a7} 
\end{eqnarray} 
is fulfilled, the collision will always be characterized 
by momentum transfers to the atom 
which are very large on the scale of the latter. 
This enables one to greatly simplify and, as we shall see below, 
also to improve the description of the excitation of the ion. 
In what follows we shall assume that the condition (\ref{a7}) is fulfilled.  

Let us first consider the elastic atomic mode. 
In this mode, because of the rapidly oscillating exponent  
$ \exp(i {\bf q}_a \cdot \mbox{\boldmath$\xi$}) $,  
the atomic elastic form-factors can be approximated by 
$ F_{A}^{0} = Z_A  - \langle u_0 | \exp(i {\bf q}_a \cdot %
\mbox{\boldmath$\xi$}) | u_0 \rangle \approx Z_A $ 
and $ F_{A}^{l} = - \langle u_0 | \alpha_l \, %
\exp(i {\bf q}_a \cdot \mbox{\boldmath$\xi$}) | u_0 \rangle \approx 0$, 
respectively. Then, taking into account (\ref{a1}), we obtain 
\begin{eqnarray} 
\frac{d^2 \sigma_{0 \rightarrow n}^{0 \rightarrow 0}}{d^2{\bf q}_{\perp}}  
= \frac{4 Z_A^2}{ v^2 } 
\frac{ \left| \left\langle \varphi_f \left| \exp(i {\bf q}_0 \cdot {\bf r} ) 
\left( 1 - \frac{v}{c} \alpha_3 \right)  
\right| \varphi_i \right\rangle \right|^2 }
{\left( {\bf q}_{\perp}^2 + (\varepsilon_n - \varepsilon_0)^2/(v^2 \gamma^2)  \right)^2 }.    
\label{a8}
\end{eqnarray} 

Let us now turn to the inelastic atomic mode. 
Because the momentum transferred to the atom is large, 
the inelastic form-factors of the atom (\ref{a3}) 
may be not small only provided the final state 
$u_m$ is a continuum state in which the momentum of the 
outgoing atomic electron ${\bf k}_a$ is approximately equal 
to the momentum transfer ${\bf q}_a$. Only then the 
rapidly oscillating exponent  
$ \exp(i {\bf q}_a \cdot \mbox{\boldmath$\xi$}) $ in 
the integrands of Eqs. (\ref{a3}) 
could be compensated by a similar factor 
$ \exp( i {\bf k}_a \cdot \mbox{\boldmath$\xi$}) $ 
which is contained in the state $u_{{\bf k}_a}$ ($ = u_m$). 

Further, since in order to "balance" the large ${\bf q}_a$ 
the momentum of the emitted electron has to be 
as large (and thus $k_a \gg Z_A$), 
its motion can be to a good approximation 
described by replacing the Coulomb atomic wave  
$u_{{\bf k}_a}$ by a corresponding plane wave. 
Physically it means that the electron of the atom 
in the collision process can be treated as quasi-free with respect 
to the nucleus of the atom and that  
the initial bound nature of the electron is  
reflected in the process merely by the Compton 
profile of the state $u_0$, which appears in 
the consideration in a natural way 
once the state $u_{{\bf k}_a}$ has been 
approximated by a plane wave. 

Taking into account also the consideration of the elastic atomic mode
we arrive at a rather simple picture of the collision process.   
In this picture, due to very large momentum transfers involved,  
the excitation of the ion is produced by the independent (incoherent)  
actions of the two quasi-free particles -- the nucleus and the electron --  
constituting initially the atom.  

One has, however, to keep in mind the following. 
Large momentum transfers are caused by a very 
large difference between the initial and final energies of 
the electron of the ion. This difference is in turn 
the consequence of a very high charge of the nucleus of the ion.  
Because of the latter the field produced by the nucleus of the ion  
in the collision can be so strong that it may determine 
the character of the motion of the electron of the atom 
in the excitation process. 

In order to see this let us make some simple estimates. 
This is convenient to do in the rest frame of the ion. 
The range of impact parameters $b_e$ of the incident atomic electron 
with respect to the nucleus of the ion, which are typical for 
the excitation process, can be roughly estimated by comparing 
the collision time $T \sim b_e / \gamma_e v_e$  
($v_e \approx v$ is the velocity of the atomic electron with respect 
to the nucleus of the ion and $\gamma_e \approx \gamma$ 
the corresponding Lorentz factor) with the transition time 
$\tau \sim Z_I^{-2}$ of the electron of the ion. 
Since for the excitation to proceed effectively 
one needs $T \stackrel{<}{\sim} \tau$, 
we obtain $ b_e \stackrel{<}{\sim} \gamma_e v_e/Z_I^2$. 
On the other hand, by comparing the force, which acts  
between the atomic electron and the nucleus of the atom, 
with the force exerted on this electron by the nucleus of the ion,  
we see that the latter will be the dominant one 
when the atomic electron enters the sphere, which is centered on 
the nucleus of the ion and has the radius $R_I = R_A \sqrt{Z_I/Z_A}$ 
(where $ R_A \simeq 1/Z_A$ is the size of the atom).  
Therefore, provided the inequality $b_e \ll R_I$, which can be written in the form  
\begin{eqnarray} 
\sqrt{ \frac{ Z_I }{ Z_A } } \, \frac{ Z_I^{2} }{ Z_A } \gg \gamma v, 
\label{ion-main-field}
\end{eqnarray}
is fulfilled the motion of the atomic electron 
in the projectile-electron excitation process   
will be predominantly governed by the field 
of the nucleus of the ion. 
Comparing (\ref{ion-main-field}) and (\ref{a7}) 
and taking into account that 
$ \sqrt{ Z_I / Z_A} > 1$ (or $ \gg 1$), 
we see that in collisions with momentum transfers, 
which are very large on the scale of the atom,  
the condition (\ref{ion-main-field}) 
is indeed always fulfilled. 

Provided the condition (\ref{a7}) is fulfilled, 
the main interaction acting on the nucleus of the atom 
in the collision process is of course also 
due to the field of the nucleus of the ion. 

A simple estimate for the magnitude 
of the effect of the field of the ionic nucleus  
on the motion of the electrons and 
the nucleus of the atom in the process 
of excitation can be obtained 
in the following way. Assume that 
there is a particle with a charge $z$ and mass $m$ 
which is incident with a velocity $v$ on the nucleus $Z_I$. 
The change in the momentum of this particle caused by the 
field of $Z_I$ is roughly given by $q \sim Z_I z/(bv)$, where 
$b$ is the impact parameter. For the problem of excitation 
the typical impact parameters are of the order of $1/Z_I$ or larger. 
One can estimate the effect of the field 
by using the ratio $\varsigma = |q|/p_i$, where  
$p_i = m \gamma v$ is the initial momentum of the incident 
particle and, thus, 
\begin{eqnarray} 
\varsigma = \frac{ |z| }{ m \gamma } \frac{ Z_I^2 }{ v^2 }. 
\label{effect}
\end{eqnarray}
From this estimate it is obvious that for the impact energies of interest 
the field of the nucleus of the ion does not really affect 
the motion of the nucleus of the atom 
but may very strongly change the motion of the atomic electrons. 

The first order cross section (\ref{a1}),  
with which we have started our current discussion,  
of course does not take into account 
the effect of the nucleus of the ion 
on the motion of the electrons of the atom. Besides,  
this cross section also does not account for the exchange  
effect of the atomic and ionic electrons. We, however, 
have already seen that in very asymmetric collisions 
the general two-center problem of excitation can be  
reduced to a single-center one 
in which only one center of force -- the nucleus 
of the ion -- is effectively present. 
Therefore, one can improve our description 
of the excitation process in collisions with large momentum transfers  
if, instead of regarding the atomic electrons 
as (quasi-) free, we would treat these electrons, 
both their initial and final states,  
as moving in the field of the nucleus of the ion  
and, besides, would take into account the exchange effect.   

In such an approach the cross section for the excitation 
of a heavy highly charged ion in collisions with an atom is 
given by the incoherent addition of the cross sections  
for excitation by the impacts of the atomic nucleus 
and electrons. This means that, if one would be able to 
compute the cross sections for the excitation 
by protons and electrons, one could use them 
(after averaging the cross section for excitation by electron impact 
over the Compton profile of the atomic electrons)  
for evaluating excitation cross sections in collisions with atoms. 
Therefore, in the next two sections we shall 
discuss excitation of heavy hydrogen-like ions in collisions with 
protons and electrons.  
 
\section{ Excitation in collisions with protons }   

Let us now turn to the consideration of excitation of 
heavy hydrogen-like ion in collisions with protons. 
The charge of the proton is much smaller than 
that of the highly charged nucleus of the ion. 
This means that the interaction between the proton and the electron 
of the ion in the process of excitation is much weaker 
than the interaction between the electron and the ionic nucleus 
and, hence, can be treated as a weak perturbation. 
Further, the proton mass is much heavier than that of the electron and, 
as was already mentioned, for collision energies 
of interest for the present study the influence of the field 
of the ionic nucleus on the proton motion can be ignored. 
Therefore, regarding the proton as a Dirac particle,  
one can approximate the initial and final states of the proton 
by (Dirac) plane-waves. 

In our consideration the nucleus of the ion will 
be taken as infinitely heavy representing, thus,  
just an external field. 
We shall work in the rest frame of this nucleus and 
choose its position as the origin. 
 
Taking all the above into account  
the transition amplitude 
for the excitation of the ion by proton impact 
can be written according to 
\begin{eqnarray}
S^{pr}_{fi}= - \frac{i}{c^2} \int d^4x \int d^4y \,%
j_{\mu}(x) \, D^{\mu \nu}(x-y) J_{\nu}(y). 
\label{e1}
\end{eqnarray}
Here, $j_{\mu}(x)$ and $J_{\nu}(y)$ ($\mu, \nu=0,1,2,3$) 
are the electromagnetic transition 4-currents 
generated by the electron of the ion 
at a space-time point $x$ and by the proton 
at a space-time point $y$, respectively,
and $D^{\mu \nu}(x-y)$ is the propagator 
of the electromagnetic field 
which transmits the  
interaction between these particles. 
The contravariant $a^{\mu}$ and 
covariant $a_{\mu}$ $4$-vectors are given by
$a^{\mu}=(a^0,{\bf a})$ and $a_{\mu}=(a^0,-{\bf a})$. 
The metric tensor $g_{\mu \nu}$
of the four-dimensional space is defined by 
$g_{00} = -g_{11}=-g_{22}=-g_{33}=1 $ 
and $g_{\mu \nu}=0 $ for $\mu \neq \nu$. 
In (\ref{e1}) the summation  
over the repeated greek indices is implied.       

The transition currents of the electron and proton 
are given by  
\begin{eqnarray}
j_{\mu}(x) = - c \overline{ \psi}_f({\bf r},t) \, \gamma_{\mu}   %
\psi_i({\bf r},t) 
\label{e2} 
\end{eqnarray}
and 
\begin{eqnarray}
J_{\mu}(y) = c \overline{ \Psi}_f({\bf R},T) \, \gamma_{\mu}   %
\Psi_i({\bf R},T), 
\label{e3} 
\end{eqnarray}
respectively, where $\gamma_{\mu}$ are the gamma-matrices. 
In Eq.(\ref{e3}) 
the vector ${\bf r}_e$ denotes the electron coordinates, 
$\psi_i({\bf r}_e,t) = \varphi_i({\bf r}_e) \exp(- i \varepsilon_i t) $ 
and $ \psi_f({\bf r}_e,t) = \varphi_f({\bf r}_e) \exp(- i \varepsilon_f t)$ 
are the initial and final states of the electron 
with total energies $\varepsilon_i$ and $\varepsilon_f$,  respectively. 
These states describe the motion of the electron in the field of the nucleus of the ion.  

Further, ${\bf R}$ are the coordinates of the proton, 
$\Psi_i({\bf R},T) = 
\phi_i({\bf R}) \exp(- i E_i T) $ and 
$\Psi_f({\bf R},T) = \phi_f({\bf R}) \exp(- i E_f T)$ 
are the initial and final states of the proton  
with  corresponding total energies $E_i$ and $E_f$.  
These states describe a free proton 
with a given value of spin projection. 

By applying the Fourier transformation to the currents and 
the photon propagator in the integrand of (\ref{e1}) 
the transition amplitude can be rewritten 
in a more convenient form 
\begin{eqnarray}
S^{pr}_{fi}= - \frac{4 \pi i}{c^3} \int d^4q  
\, \widetilde{j}_{\mu}(q) \frac{1}{q^2 + i0 } %
\widetilde{J}^{\mu}(-q),  
\label{e4}
\end{eqnarray}
where 
\begin{eqnarray}
\widetilde{j}_{\mu}(q) &=& \frac{1}{4 \pi^2 } \int d^4x 
\, j_{\mu}(x) \exp(-i q x) 
\nonumber \\ 
&=& 
\frac{ 1 }{ 2\pi } \delta\left( q_0 + (\varepsilon_i-\varepsilon_f)/c \right) \int d^3 {\bf r} 
\overline{ \varphi }_f({\bf r})  \gamma_{\mu} 
\exp(i {\bf q} \cdot {\bf r} ) \varphi_i({\bf r}) 
\nonumber \\ 
\widetilde{J}_{\mu}(-q) &=& \frac{1}{4 \pi^2 } \int d^4y 
\, J_{\mu}(y) \exp( i q y ) 
\nonumber \\ 
&=& 
\frac{ 1 }{ 2\pi } \delta\left( q_0 - (E_i-E_f)/c \right) \int d^3 {\bf R} 
\overline{ \phi }_f({\bf R})  \gamma_{\mu} 
\exp(- i {\bf q} \cdot {\bf R} ) \phi_i({\bf R}).  
\label{e5}
\end{eqnarray} 

Due to the relatively large mass of the proton 
the change in its initial momentum caused by 
the collision is much smaller than the initial 
momentum itself. As a result, the proton not 
only moves in the collision practically along 
a straight line but also, as one can easily show, 
the change in the direction of its spin is very unlikely. 
Taking this into account, assuming for definiteness 
that the proton moves initially along the $z$-axis 
and using the explicit form of the Dirac plane-wave states 
for the proton its current in Eqs.(\ref{e5}) 
can be greatly simplified resulting in 
\begin{eqnarray}
\widetilde{J}_{0}(-q) &=&  
\frac{ 1 }{ 2\pi } \delta\left( q_0 - (E_i-E_f)/c \right) 
\delta^{(3)}\left({\bf P}_i - {\bf q} - {\bf P}_f  \right) 
\nonumber \\ 
\widetilde{J}_{3}(-q) &=&  
\frac{ v }{ 2\pi c} \delta\left( q_0 - (E_i-E_f)/c \right)  
\delta^{(3)}\left({\bf P}_i - {\bf q} - {\bf P}_f  \right) 
\nonumber \\ 
\widetilde{J}_{1}(-q) &=& 0 
\nonumber \\ 
\widetilde{J}_{2}(-q) &=& 0.  
\label{e6} 
\end{eqnarray} 
In the above equations ${\bf P}_i = (0,0,P_i)$ and ${\bf P}_f$ are 
the initial and final momenta of the proton, respectively, 
($|{\bf P}_i - {\bf P}_f | \ll |{\bf P}_i| $) 
and $v$ is the proton velocity with respect to the nucleus, 
which to excellent accuracy remains a constant in the collision. 

Using Eqs. (\ref{e4}) and (\ref{e6}) and the expression for 
the electron transition current from Eq. (\ref{e5}) 
the transition amplitude is obtained to be 
\begin{eqnarray}
S^{pr}_{fi} &=& \frac{i}{ \pi } 
\frac{  \delta(\varepsilon_i + E_i - \varepsilon_f - E_f ) }  
{ {\bf q}^2 - (\varepsilon_f - \varepsilon_i)^2 / c^2 }  
\int d^3 {\bf r} \varphi^{\dagger}_f({\bf r})   
\exp(i {\bf q} \cdot {\bf r} )  
\left( 1 - \frac{v}{c} \alpha_3 \right) \varphi_i({\bf r})  
\nonumber \\ 
 & = &  \frac{ i }{ \pi } \delta(\varepsilon_i + E_i - \varepsilon_f - E_f ) %
 \frac{ \left\langle \varphi_f \left| \exp(i {\bf q} \cdot {\bf r} ) 
\left( 1 - \frac{v}{c} \alpha_3 \right)  
\right| \varphi_i \right\rangle }
{ {\bf q}^2 - (\varepsilon_f - \varepsilon_i)^2/c^2 }, 
\label{e7}
\end{eqnarray}
where ${\bf q} = {\bf P}_i - {\bf P}_f$ is the momentum transfer 
to the ion, $\alpha_3$ is the Dirac matrix 
and the delta-function expresses 
the energy conservation in the collision.  
  
Using the well known procedure in order to obtain  
the cross section from the transition amplitude 
for the excitation cross section differential in the momentum 
transfer we get 
\begin{eqnarray}
\frac{d^3 \sigma_{fi}}{d {\bf q}^3} = \frac{4}{ v } 
\frac{ \left| \left\langle \varphi_f \left| \exp(i {\bf q} \cdot {\bf r} ) 
\left( 1 - \frac{v}{c} \alpha_3 \right)  
\right| \varphi_i \right\rangle \right|^2 }
{ \left( {\bf q}^2 - (\varepsilon_f - \varepsilon_i)^2/c^2 \right)^2 } 
\delta(\varepsilon_i + E_i - \varepsilon_f - E_f ).  
\label{e8}
\end{eqnarray}
In the above consideration we have already used the fact that 
the change in the momentum of the proton caused 
by the collision is very small compared to its initial value. 
Using this fact again one can show that, to an excellent accuracy,  
the change in the proton energy is very simply related to 
the $z$-component, $q_z$, of the momentum transfer vector 
${\bf q}$: $ E_i - E_f = v q_z$. This enables us to integrate 
the cross section (\ref{e8}) over $q_z$ and obtain 
\begin{eqnarray}
\frac{d^2 \sigma_{fi}}{d {\bf q}_{\perp}^2} = \frac{4}{ v^2 } 
\frac{ \left| \left\langle \varphi_f \left| \exp(i {\bf q}_0 \cdot {\bf r} ) 
\left( 1 - \frac{v}{c} \alpha_3 \right)  
\right| \varphi_i \right\rangle \right|^2 }
{ \left( {\bf q}_0^2 - (\varepsilon_f - \varepsilon_i)^2/c^2 \right)^2 },   
\label{e9}
\end{eqnarray}
where 
\begin{eqnarray}
{\bf q}_0 = \left( {\bf q}_{\perp}, q_{min} \right) 
\label{e10}
\end{eqnarray}
with ${\bf q}_{\perp}$ being the transverse part of the momentum 
transfer (${\bf q}_{\perp} \cdot {\bf v} = 0$) and   
\begin{eqnarray} 
q_{min}= \frac{ \varepsilon_f - \varepsilon_i }{ v }    
\label{e11} 
\end{eqnarray}
is the minimum momentum transfer in the collision.    
It is not difficult to see that the only difference between 
the cross section (\ref{a8}), obtained in the previous section, 
and the cross section (\ref{e9}) and is that 
the latter was derived by assuming $Z_A=1$.  

The initial and final bound states of the electron 
in (\ref{e9}) are given by 
\begin{eqnarray}
\varphi_i ({\bf r_e})  =  
\left(
\begin{array}{cc}
 g_{{n_i}\kappa_i} (r_e)  & \chi_{\kappa_i}^{\mu_i}  ({\bf \hat{r}_e} )\\
i f_{{n_i}\kappa_i}(r_e)  & \chi_{-\kappa_i}^{\mu_i} ({\bf \hat{r}_e }) \\
\end{array}
\right)
\label{e12}
\end{eqnarray}
and
\begin{eqnarray}
\varphi_f ({\bf r_e})  =  
\left(
\begin{array}{cc}
 g_{{n_f}\kappa_f} (r_e)  & \chi_{\kappa_f}^{\mu_f}  ({\bf \hat{r}_e } )\\
i f_{{n_f}\kappa_f}(r_e)  & \chi_{-\kappa_f}^{\mu_f} ({\bf \hat{r}_e }) \\
\end{array}
\right), 
\label{e13}
\end{eqnarray} 
respectively. In Eqs. (\ref{e12}) and (\ref{e13}) 
$g_{n\kappa}$ ($f_{n\kappa}$) are the large (small) 
components of the radial Dirac-Coulomb states 
of the electron in the field of the nucleus 
with a charge $Z_I$. 

If we denote $ \zeta = Z_I \alpha $, 
where $\alpha = e^2/{\hbar c} = 1/137.04$ is the fine structure constant, 
and $ \Lambda = \sqrt{\kappa^2 - \zeta^2}$, 
then the energies and the radial wave functions 
of the bound states are given, respectively, by (see e.g. \cite{rose} ) 
\begin{eqnarray}  
\varepsilon = m c^2 \left[ 1+\left(\frac{\zeta}{n' + \Lambda}\right)^2 \right]^{-\frac{1}{2}} 
\label{e14} 
\end{eqnarray} 
and 
\begin{eqnarray}
\left\{
\begin{array}{c}
 g_{n\kappa} \\
 f_{n\kappa} 
\end{array}
\right\}& = &
\pm (1 \pm  \varepsilon/mc^2)^{\frac{1}{2}} 
\frac{\sqrt{2} k^{\frac{5}{2}} \lambdabar_e } { \Gamma(2\Lambda+1)}  
\left( \frac{\Gamma(2\Lambda + n'+1)}{n'! \zeta %
(\zeta-\kappa k \lambdabar_e)} \right)^{1/2} (2 k r )^{\Lambda-1} e^{-k r} 
\nonumber \\ 
& \times & [ \mp n' _1 F_1(-n' + 1 , 2\Lambda+1 ; 2 k r) - 
(\kappa-\frac{\zeta}{k\lambdabar_e})_1 F_1(-n' , 2\Lambda+1 ; 2 k r) ]. 
\label{e15} 
\end{eqnarray}
Here, $\Gamma(x)$ and $_1 F_1(a,b;z)$ are the gamma function and confluent 
hypergeometric function \cite{Ab-St}, respectively, 
$\lambdabar_e = \hbar/mc $ is the electron Compton wavelength,  
$k = \frac{\zeta}{\lambdabar_e} [\zeta^2 + (n'+\Lambda)^2]^{-\frac{1}{2}}$,  
$n = n' + |\kappa|$ is the principal quantum number 
and the quantity $\kappa$ is related to 
the orbital momentum $l$ and the total angular momentum $j$ by 
\begin{eqnarray}
l = \left\{
\begin{array}{cc}
\kappa, & \mbox{if} \qquad \kappa > 0 \\
-\kappa -1, & \mbox{if} \qquad  \kappa < 0 \\
\end{array} \right.
\mbox{and} \qquad j = |\kappa| - \frac{1}{2}. 
\label{e16} 
\end{eqnarray}
Further, $\chi_\kappa^\mu$ are the normalized 
spin-angular functions (see e.g. \cite{rose}) 
which read 
\begin{eqnarray}
\chi_\kappa^\mu ({\bf \hat{r}_e} )  =  \sum_{m_l} \left(
\begin{array}{cc|c}
 l   & \frac{1}{2} & j \\
 m_l &  \mu -m_l   &  \mu \\
\end{array}
\right) Y^*_{l m_l}({\bf \hat{r}_e }) \chi^{\mu -m_l }_{\frac{1}{2}},  
\end{eqnarray}
where 
\begin{eqnarray} 
\chi^{\frac{1}{2}}_{\frac{1}{2}} =  
\left( \begin{array}{c}
 1  \\
 0  \\ 
\end{array} \right) 
\nonumber \\ 
\chi^{-\frac{1}{2}}_{\frac{1}{2}} =  
\left( \begin{array}{c}
 0  \\
 1  \\ 
\end{array} \right) 
\end{eqnarray}
are the Pauli spinors and 
$Y_{l m_l}$ are the spherical harmonics.  

\section{ Excitation by electron impact }

Like in the case of collisions with protons, 
the interaction between the incident electron 
and the electron of the ion is comparatively very weak. 
Therefore, this interaction can be treated 
as arising due to just single-photon exchange 
between the electrons. 

However, there are in general two important 
differences between the excitation 
of a highly charged ion by proton and electron impacts. 
First, the mass of an electron is much lighter than 
that of a proton. As a result, in contrast to the proton case, 
the motion of the incident (and scattered) electron 
can be very substantially distorted by its interaction 
with the nucleus of the ion.  
Second, since the electrons are indistinguishable, 
the exchange effect has to be taken into account. 

The first point can be addressed by describing 
not only the bound but also the continuum electron as 
moving in the Coulomb field of the nucleus of the ion. 
The second point leads to the necessity to include 
an additional first order diagram (the so called exchange diagram) 
into the theoretical treatment of electron-impact excitation.   

Taking all the above into account the transition amplitude 
for electron-impact excitation is given by 
\begin{eqnarray}
S^{tot}_{fi}= S^{dir}_{fi} + S^{exc}_{fi},  
\label{ee1}
\end{eqnarray}
where $S^{dir}_{fi}$ and $S^{exc}_{fi}$ 
are the direct and exchange contributions, 
respectively, to the total transition amplitude. 

Similarly to the case of proton impact 
for the direct contribution one can obtain   
\begin{eqnarray}
S^{dir}_{fi} = - \frac{4 \pi i}{c^3} \int d^4q  
\, \widetilde{j}^{dir}_{\mu}(q) \frac{1}{q^2 + i0 } %
\widetilde{J}_{dir}^{\mu}(-q),  
\label{ee2}
\end{eqnarray}
where 
\begin{eqnarray}
\widetilde{j}^{dir}_{\mu}(q) = 
\frac{ 1 }{ 2\pi } \delta\left( q_0 + (\varepsilon_i-\varepsilon_f)/c \right) \int d^3 {\bf r} 
\overline{ \varphi }_f({\bf r})  \gamma_{\mu} 
\exp(i {\bf q} \cdot {\bf r} ) \varphi_i({\bf r}) 
\label{ee3}
\end{eqnarray} 
describes the current generated by the electron in its bound-bound transition 
in the ion and 
\begin{eqnarray}
\widetilde{J}^{dir}_{\mu}(-q) =  
\frac{ 1 }{ 2\pi } \delta\left( q_0 - (E_i-E_f)/c \right) \int d^3 {\bf r} 
\overline{ \varphi }_{{\bf p}_f}({\bf r})  \gamma_{\mu} 
\exp(- i {\bf q} \cdot {\bf r} ) \varphi_{{\bf p}_i}({\bf r})   
\label{ee4}
\end{eqnarray} 
represent the current generated 
by the incident and scattered electron with asymptotic 
momenta ${\bf p}_i$ and ${\bf p}_f$, respectively,  
(continuum-continuum transition). 

The exchange part of the transition amplitude reads 
\begin{eqnarray}
S^{exc}_{fi} = + \frac{4 \pi i}{c^3} \int d^4q  
\, \widetilde{j}^{exc}_{\mu}(q) \frac{1}{q^2 + i0 } %
\widetilde{J}_{exc}^{\mu}(-q).   
\label{ee5} 
\end{eqnarray}
In this expression 
\begin{eqnarray}
\widetilde{j}^{exc}_{\mu}(q) =   
\frac{ 1 }{ 2\pi } \delta\left( q_0 + (\varepsilon_i - E_f)/c \right) \int d^3 {\bf r} 
\overline{ \varphi }_{{\bf p}_f}({\bf r})  \gamma_{\mu} 
\exp(i {\bf q} \cdot {\bf r} ) \varphi_i({\bf r}) 
\label{ee6}
\end{eqnarray} 
describes the current generated by the electron, 
which was initially bound in the ion and emitted 
during the collision having asymptotically 
a momentum ${\bf p}_f$ (bound-continuum transition).  
Further, 
\begin{eqnarray}
\widetilde{J}^{exc}_{\mu}(-q) =  
\frac{ 1 }{ 2\pi } \delta\left( q_0 - (E_i - \varepsilon_f)/c \right) \int d^3 {\bf r} 
\overline{ \varphi }_f({\bf r})  \gamma_{\mu} 
\exp(- i {\bf q} \cdot {\bf r} ) \varphi_{{\bf p}_i}({\bf r})   
\label{ee7} 
\end{eqnarray} 
represent the current generated 
by the electron, which was initially incident on the ion 
with an asymptotic momentum ${\bf p}_i$  
and become bound in the collision (continuum-bound transition). 

In the above expressions the form of the bound states 
is given, as before, by Eqs. (\ref{e12})-(\ref{e16}). 
The continuum states (the incident and scattered electron(s))  
are described using the corresponding Dirac wave function 
in the Coulomb field of the nucleus of the ion.  

Namely, for the incident electron 
which propagates initially in the positive z-axis 
direction $({\bf p}_i =(0,0,p_i))$ 
and has spin projection $m_s$ one has 
\begin{eqnarray}
\label{eq6}
\psi_i^{(m_s)} ({\bf r} , t)  = & e^{-i E_i t }  %
 \sqrt{\frac{\pi c^2}{2 p_i E_i  }}
\sum_{\kappa_i}  i^l \sqrt{4 \pi (2 l +1)}
\left(
\begin{array}{cc|c}
 l   & \frac{1}{2} & j_i \\
 0 & m_s & m_s\\
\end{array}
\right) 
\nonumber \\ 
& \times  e^{ i \Delta_{\kappa_i}}
\left(
\begin{array}{cc}
 g_{{E_i} \kappa_i} (r)  & \chi_{\kappa_i}^{m_s}  ({\bf \hat{r}} )\\
i f_{{E_i} \kappa_i}(r)  & \chi_{-\kappa_i}^{m_s} ({\bf \hat{r} }) \\
\end{array}
\right). 
\end{eqnarray}
The state of the scattered electron, which asymptotically 
has a momentum ${\bf p}_f$ and spin projection $m_s$, reads 
\begin{eqnarray}
\label{eq7}
\psi_f^{(m_s)} ({\bf r} , t)  = & e^{-i E_f t }  4 \pi \sqrt{\frac{\pi c^2}{2 p_f E_f  }} 
\sum_{\kappa_{f} \mu_f} i^l
\left(
\begin{array}{cc|c}
 l   & \frac{1}{2} & j_f \\
 m_l  & m_s & \mu_f \\
\end{array}
\right) 
\nonumber \\ 
& \times  e^{- i \Delta_{\kappa_f}} Y^*_{lm_l}({\bf \hat{p}_f })
\left(
\begin{array}{cc}
 g_{{E_f}\kappa_f} (r)  & \chi_{\kappa_f}^{\mu_f}  ({\bf \hat{r} } )\\
i f_{{E_f}\kappa_f}(r)  & \chi_{-\kappa_f}^{\mu_f} ({\bf \hat{r} }) \\
\end{array}
\right). 
\end{eqnarray} 

The radial wave functions $g_{E\kappa} (r)$ and $f_{E\kappa} (r)$ are given by 
\begin{eqnarray}
\left\{
\begin{array}{c}
 g_{E\kappa} \\
 f_{E\kappa} 
\end{array}
\right\} & = &
\pm (E\pm mc^2)^{\frac{1}{2}} 
\frac{2 p^{\frac{1}{2}}} {c \pi^\frac{1}{2}} (2 p r )^{\Lambda-1} e^{\pi \eta/2} 
\frac{|\Gamma(\Lambda + i\eta)|}{\Gamma(2\Lambda+1)} 
\nonumber \\ 
&& \times \left\{
\begin{array}{c}
 \Re \\
 \Im 
\end{array}
\right\} e^{-i p r} e^{i \delta_{\kappa}} (\Lambda + i \eta) %
 _1 F_1(\Lambda + 1 + i\eta; 2\Lambda +1 , 2 i p r) 
\label{continuum} 
\end{eqnarray}
where $\eta = \frac{Z_I E  }{ p  c^2  }$ 
is the Sommerfeld parameter, the Coulomb phase 
shift $\delta_\kappa $ is defined by the relation 
$$ e^{2 i \delta_\kappa}  = \frac{-\kappa + i c^2 \eta /E}{\Lambda+ i \eta} $$
and $\Delta_\kappa = \delta_\kappa -arg \Gamma(\Lambda+i \eta) -\frac{\pi}{2} \Lambda $ 
(see \cite{rose}). In expression (\ref{continuum}) the notations $\Re$  and $\Im$ mean 
that one has to take the real or the imaginary  
part, respectively, of its second line. 
 
\section{ Some numerical results and discussion }


\begin{figure}[t]
\vspace{0.25cm} 
\begin{center}
\includegraphics[width=0.55\textwidth]{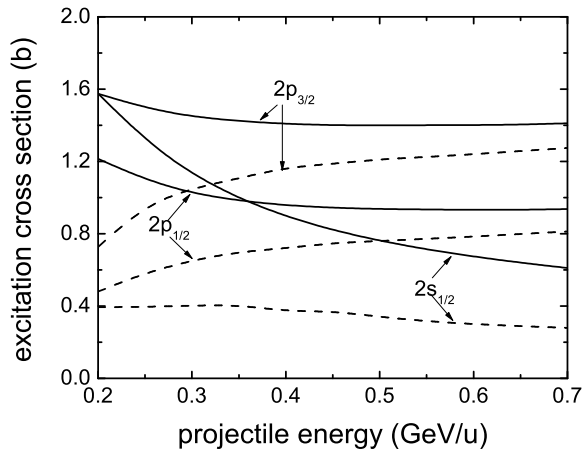} 
\end{center}
\caption{ \footnotesize{ 
Cross sections for excitation of Bi$^{82+}$(1s) projectiles  
into the $L$-shell in collisions with hydrogen 
in the inelastic target mode. Solid curves: the results 
for the excitation by electron impact.  
Dash curves: the results of the first-order 
perturbation theory in the ion-atom interaction. 
For more explanation see the text. } } 
\label{inel_mode}  
\end{figure} 

Figure \ref{inel_mode} shows the contribution of the inelastic target 
mode to the cross sections for excitation of Bi$^{82+}$(1s) projectiles 
into the $L$-shell occurring in collisions with atomic hydrogen: 
Bi$^{82+}$(1s) $+ $ H(1s) $\to$ Bi$^{82+}$($n=2,j$) $+$ p$^+$ + e$^-$. 
The figure contains two sets of the theoretical results. One of them, 
depicted by dash curves, was obtained by using the first order of perturbation 
theory in the projectile-target interaction. 
The other one, displayed by solid curves, was calculated by treating 
the target electron in the inelastic mode as quasi-free with respect 
to the target nucleus but taking into account the distortion of its motion 
by the field of the nucleus of the projectile and also the exchange effect between 
the electrons of the target and projectile (see section 4).  
  
As it follows from figure \ref{inel_mode}, there is a large 
difference between these two sets of the cross sections 
with the distorted-wave model yielding substantially 
higher cross sections. This is especially obvious 
in the case of the excitation of 
the $1s_{1/2}$--$2s_{1/2}$ transition where 
even at an impact energy as high as $\simeq 700$ MeV/u 
the difference still reaches a factor of $2$. 

The origin of this difference may lie 
in the strong attraction between the electron of the atom 
and the nucleus of the ion which increases the probability for 
the atomic electron to come closer to the electron of the ion. 
Such a "focusing" could be especially effective 
namely in the excitation of the $1s_{1/2}$--$2s_{1/2}$ 
transition since the latter needs small 
impact parameters in order to proceed.  

When the impact energy increases the distortion of 
the motion of the incident electron in general weakens. 
However, according to the results shown in figure \ref{inel_mode} 
the distortion presents a substantial factor for all energies  
from the range considered and the difference in the cross sections 
remains quite substantial. 

\vspace{0.5cm} 

\begin{figure}[t]
\vspace{0.25cm} 
\begin{center}
\includegraphics[width=0.55\textwidth]{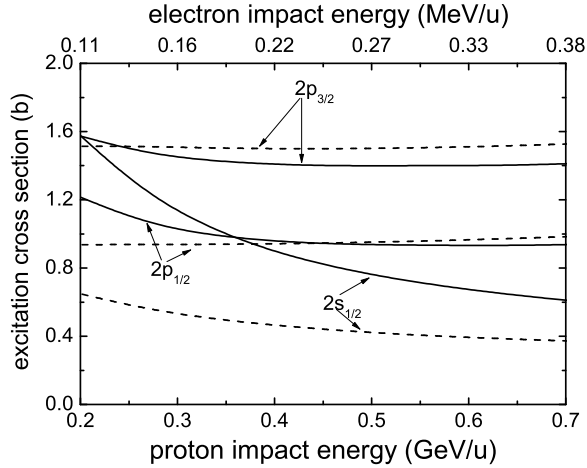} 
\end{center}
\caption{ \footnotesize{ 
Cross sections for excitation of Bi$^{82+}$(1s) 
ions into the $L$-shell by the impact  
of equivelocity electrons (solid curves) and protons (dash curves).  } } 
\label{elect-prot}  
\end{figure} 

In figure \ref{elect-prot} we compare  
the cross sections for excitation 
of Bi$^{82+}$(1s) ions by the impact of the proton and electron  
which have (initially) equal velocities with respect to the ion.  
Note that according to the present approach 
these cross sections also correspond to the contributions to the excitation 
by the elastic (proton) and inelastic (electron) target modes 
in collisions with atomic hydrogen. It is seen that in the interval 
of collision velocities considered in the figure 
the results for excitation by electrons and protons are rather different.  
In particular, despite the incident protons possess kinetic energy, which is 
by three orders of magnitude larger than that of the equivelocity electrons,  
it turns out that the latter ones can be more effective 
in exciting the $1s_{1/2}$--$2s_{1/2}$ and $1s_{1/2}$--$2p_{1/2}$ 
transitions.  

The "superiority" of the electrons in the excitation process  
is most substantial for the $1s_{1/2}$--$2s_{1/2}$ transition for which 
the electron is more efficient by a factor of $2.5$ 
at the lower boundary of the velocity interval.  
The origin of this difference most likely lies 
in a very strong distortion of the motion of 
the incident electron by the field 
of the nucleus of the ion. While the heavy proton 
moves in the collision practically undeflected 
the strong attraction of the light incident electron 
by the field of the ion increases the probability 
for this electron to come closer to the 
location of the bound electron. Since 
the $1s_{1/2}$-- $2s_{1/2}$ transition, as a non-dipole one, 
occurs at small impact parameters, the "focusing" 
of the incident electron by the field of the nucleus 
may increase the chances for the excitation.     



\section{ Conclusions }

We have considered excitation of highly charged hydrogen-like 
ions in relativistic collisions with light 
atoms in which the momentum transfer to the atom 
is very large on the typical atomic scale. 
In the process of excitation in such collisions 
the nucleus and the electrons of the atom 
behave as quasi-free particles 
with respect to each other and 
it is the field of the nucleus of the ion which 
is the main force acting on them.
Therefore, excitation of the ion essentially proceeds via  
two distinct reaction pathways, which involve  
the collision of the electron of the ion either  
with the atomic nucleus or with the atomic electrons, 
whose contributions add up incoherently 
in the cross section.  

Since the nucleus of the atom is much heavier 
than the electron, 
its motion remains practically not distorted 
by the interaction with the field 
of the nucleus of the ion. Because of that and 
also due to the condition $Z_I \gg Z_A$ 
the contribution to the excitation of the ion,  
caused by the interaction with the nucleus of the atom, 
can be evaluated already within the first 
order perturbation theory in the interaction 
between the nucleus of the atom 
and the projectile-ion.  

Contrary to this, the field of the nucleus of the projectile 
in collisions with large momentum transfer 
has a crucial impact on the motion of the electrons of the atom. 
Therefore, for a proper description of the excitation by atomic 
electrons one needs to take into account the distortion 
of their states both in the initial and final reaction channels. 
Besides, since there is in general a noticeable overlap 
between the phase spaces of the electron of the ion and those of the atom 
the exchange effect between them has to be taken into account.  

\section{Acknowledgement} 

A.B.V. acknowledges the support from the Extreme Matter Institute EMMI.

\end{document}